\documentclass{emulateapj}
\def \mlb{\hbox{$h{\rm M}_\odot/{\rm L}_{{\rm B}\odot}$}}
\def \mlr{\hbox{$h{\rm M}_\odot/{\rm L}_{{\rm R}\odot}$}}
\def \mlg{\hbox{$h{\rm M}_\odot/{\rm L}_{{\rm g'}\odot}$}}
\def \lsun{\hbox{${\rm L}_\odot$}}
\def \lbsun{\hbox{${\rm L}_{B\odot}$}}
\def \msun{\hbox{${\rm M}_\odot$}}

\lefthead{Hoekstra et al.}
\righthead{Virial masses and the baryon fraction in galaxies}

\begin{document}

\title{Virial masses and the baryon fraction in
galaxies\footnotemark[$\dagger$]}

\footnotetext[$\dagger$]{Based on observations from the Canada-France-Hawaii
Telescope, which is operated by the National Research Council of
Canada, le Centre National de la Recherche Scientifique and the
University of Hawaii.}

\author{H. Hoekstra\altaffilmark{1,2,3}, B.C. Hsieh\altaffilmark{4,5},
H.K.C. Yee\altaffilmark{3}, H. Lin\altaffilmark{6}, 
M.D. Gladders\altaffilmark{7}}

\altaffiltext{1}{Department of Physics and Astronomy, University of Victoria,
Victoria, BC, V8P 5C2, Canada}
\altaffiltext{2}{CITA, University of Toronto, Toronto, Ontario M5S 3H8, Canada}
\altaffiltext{3}{Department of Astronomy \& Astrophysics,
        University of Toronto, 60 St. George Street, Toronto, Ontario
        M5S 3H8, Canada}
\altaffiltext{4}{Institute of Astronomy, National Central University,
No. 300, Jhongda Rd. Jhongli City, Taoyuan County 320, Taiwan, R.O.C.}
\altaffiltext{5}{Institute of Astrophysics \& Astronomy, Academia Sinica,
P.O. Box 23-141, Taipei 106, Taiwan, R.O.C}
\altaffiltext{6}{Fermi National Accelator Laboratory, P.O. Box 500, Batavia,
IL 60510}
\altaffiltext{7}{Carnegie Observatories, Pasadena, CA 91101, USA}

\begin{abstract}

We have measured the weak lensing signal as a function of restframe
$B$, $V$, and $R$-band luminosity for a sample of `isolated' galaxies.
These results are based on four-band photometry from the Red-Sequence
Cluster Survey, enabling us to determine photometric redshifts for a
large number of galaxies. We select a secure sample of lenses with
photometric redshifts $0.2<z<0.4$ and study the relation between the
virial mass and baryonic contents. In addition, we discuss the
implications of the derived photometric redshift distribution for
published cosmic shear studies. The virial masses are derived from a
fit to the observed lensing signal. For a galaxy with a fiducial
luminosity of $10^{10}h^{-2}$L$_{B\odot}$ we obtain a mass $M_{\rm
vir}=9.9^{+1.5}_{-1.3}\times 10^{11}$M$_\odot$.  The virial mass as a
function of luminosity is consistent with a power-law $\propto
L^{1.5}$, with similar slopes for the three filters considered here.
These findings are in excellent agreement with results from the Sloan
Digital Sky Survey and semi-analytic models of galaxy formation. We
measure the fraction of mass in stars and the baryon fraction in
galaxies by comparing the virial mass-to-light ratio to predicted
stellar mass-to-light ratios. We find that star formation is
inefficient in converting baryons into stars, with late-type galaxies
converting $\sim 33$\% and early-type galaxies converting only $\sim
14$\% of baryons into stars. Our results imply that the progenitors of
early-type galaxies must have low stellar mass fractions, suggestive
of a high formation redshift.

\end{abstract}


\section{Introduction}

Observations of rotation curves of spiral galaxies and measurements of the
velocity dispersions of stars in early-type galaxies have provided
important evidence for the existence of massive dark matter halos
around galaxies (e.g., van Albada \& Sancisi 1986). In addition, these
studies have presented evidence of tight relations between the
baryonic and dark matter components (e.g., Tully \& Fisher 1977; Faber
\& Jackson 1976). Results based on strong lensing by galaxies support
these findings (e.g., Keeton, Kochanek \& Falco 1998).

The origin of these scaling relations must be closely related to the
process of galaxy formation, but the details are still not well
understood, mainly because of the complex behaviour of the baryons.
Furthermore, on the small scales where baryons play such an important
role, the accuracy of cosmological numerical simulations is limited.
This complicates a direct comparison of models of galaxy formation to
observational data. For such applications, it would be more convenient
to have observational constraints on quantities that are robust and
easily extracted from numerical simulations. 

An obvious choice is the virial mass of the galaxy, but most
techniques for measuring mass require visible tracers of the potential,
confining the measurements to relatively small radii. Fortunately,
recent developments in weak gravitational lensing have made it possible to
probe the ensemble averaged mass distribution around galaxies out to
large projected distances. The tidal gravitational field of the dark
matter halo introduces small coherent distortions in the images of
distant background galaxies, which can be easily detected in current
large imaging surveys. We note that one can only study ensemble
averaged properties, because the weak lensing signal induced by an
individual galaxy is too small to be detected.

Since the first detection of this so-called galaxy-galaxy lensing
signal by Brainerd et al. (1996), the significance of the measurements
has improved dramatically, thanks to new wide field CCD cameras on a
number of mostly 4m class telescopes. This has allowed various groups
to image large areas of the sky, yielding the large numbers of lenses
and sources needed to measure the lensing signal. For instance,
Hoekstra et al. (2004) used 45.5 deg$^2$ of $R_C$-band imaging data
from the Red-Sequence Cluster Survey (RCS), enabling them to measure,
for the first time, the extent and flattening of galaxy dark matter
halos, providing strong support for the cold dark matter (CDM)
paradigm. However, the analysis presented in Hoekstra et al. (2004)
was based on the $R_C$-band data alone, and consequently lacked
redshift information for the individual lenses.

An obvious improvement is to obtain redshift information for the lenses
(and if possible the sources). This allows one to study the lensing
signal as a function of lens properties, most notably the luminosity.
Photometric redshifts were used by Hudson et al. (1998) to scale the
lensing signal of galaxies in the Hubble Deep Field, and by Wilson et
al. (2001) who measured the lensing signal around early-type galaxies
as a function of redshift. Smith et al. (2001) and Hoekstra et
al. (2003) used spectroscopic redshifts, but the lens samples involved
were rather small ($\sim 1000$). The Sloan Digital Sky Survey (SDSS)
combines both survey area and redshift information. Its usefulness for
galaxy-galaxy lensing was demonstrated clearly by Fischer et
al. (2000). More recently, McKay et al. (2001) used the available SDSS
redshift information to study the galaxy-galaxy lensing signal as a
function of galaxy properties (also see Guzik \& Seljak 2002; Seljak
2002; Sheldon et al. 2004).

In this paper we use a subset of the RCS data, for which photometric
redshifts have been determined using $B,V,R_C$ and $z'$ data taken
using the Canada-France-Hawaii Telescope (see Hsieh et al. 2005 for
details). The area covered by these multiwavelength data is
approximately 33.6 deg$^2$, resulting in a catalog of $1.2\times 10^6$
galaxies for which a redshift could be determined, making it one of
the largest data sets of its kind. This unique data set allows us to
measure the virial masses of galaxies as a function of their
luminosity.

This paper is structured as follows. In \S2 we briefly discuss the
data, including the photometric redshift catalog and its accuracy. The
results of some basic tests of the photometric redshifts are presented
in \S3. In \S4 we discuss the dark matter profile inferred from
numerical simulations. The measurement of the virial mass as a
function of luminosity in various filters is presented in \S5, as well
as our measurement of the baryon fraction in galaxies. Throughout the
paper we adopt a flat cosmology with $\Omega_m=0.3$,
$\Omega_\Lambda=0.7$ and a Hubble parameter $H_0=100 h$ km/s/Mpc.

\section{Data}

The Red-Sequence Cluster Survey (RCS) is a galaxy cluster survey
designed to provide a large sample of optically selected clusters of
galaxies in a large volume (see Gladders \& Yee (2005) for a detailed
discussion of the survey). To this end, 92 deg$^2$ of the sky were
imaged in both $R_C$ and $z'$ using the CFH12k camera on CFHT and the
Mosaic II camera on the CTIO Blanco telescope. This choice of filters
allows for the detection of clusters up to $z\sim 1.4$ using the cluster
red-sequence method developed by Gladders \& Yee (2000).

After completion of the original RCS survey, part of the surveyed area
was imaged in both $B$ and $V$ band using the CFHT. This additional
color information allows for a better selection of clusters at lower
redshifts. These follow-up observations cover $\sim 33.6$ deg$^2$, thus
covering $\sim 70\%$ of the CFHT fields. The data and the photometric
reduction are described in detail in Hsieh et al. (2005).

The galaxy-galaxy lensing results presented in Hoekstra et al. (2004)
were based on 45.5 deg$^2$ of $R_C$-band data alone. The addition of
$B$ and $V$ imaging data for 33.6 deg$^2$ to the existing $R_C$ and $z'$
data allow for the determination of photometric redshifts for both
lenses and sources in this subset of RCS imaging data. This enables
the study of the lensing signal as a function of the photometric
properties of the lens galaxies (i.e., color and luminosity). In this
paper we focus on this multi-color subset of the RCS.

To determine the restframe $B$, $V$ and $R$ luminosities we use
template spectra for a range in spectral types and compute the
corresponding passband corrections as a function of redshift and
galaxy color (this procedure is similar to the one described in van
Dokkum \& Franx 1996). Provided the observed filters straddle the
redshifted filter of interest, which is the case here, this procedure
yields very accurate corrections.

The CFHT $R_C$ images are used to measure the shapes of galaxies used
in the weak lensing analysis. The raw galaxy shapes are corrected for
the effects of the point spread function, as described in Hoekstra et
al. (2002a). The resulting object catalogs have been used for a range
of weak lensing studies (e.g., Hoekstra et al. 2002a, 2002b, 2002c,
2004) and we refer to these papers for a detailed discussion of the
shape measurements.

The measurements of the lensing signal caused by large scale structure
presented in Hoekstra et al. (2002a, 2002b) are very sensitive to
residual systematics. The various tests described in these papers
suggest that the systematics are well under control. In this paper we
use the shape measurements to measure the galaxy-galaxy lensing
signal, which is much less sensitive to these observational
distortions: in galaxy-galaxy lensing one measures the lensing signal
that is perpendicular to the lines connecting many lens-source pairs.
These are randomly oriented with respect to the PSF anisotropy, and
therefore residual systematics are suppressed.

\subsection{Photometric redshift distribution}

The determination of the photometric redshifts is described in detail
in Hsieh et al. (2005). The empirical quadratic polynomial fitting
technique (Connolly et al. 1995) is used to estimate redshifts for the
galaxies in the RCS data. The key component in this approach is the
creation of a training set. Spectroscopic redshifts from the CNOC2
survey (Yee et al. 2000) are matched to the corresponding objects in
the overlapping RCS fields. These data are augmented with observations
of the GOODS/HDF-N field, for which the spectroscopic redshifts have
been obtained using the Keck telescope (Wirth et al. 2004; Cowie et
al. 2004), and the photometry is from the ground-based Hawai'i HDF-N
data obtained with the Subaru telescope (Capak et al. 2004).  This
results in a final training set that includes 4,924 objects covering a
large range in redshifts. To minimize the fitting errors arising from
different galaxy types, Hsieh et al. (2005) used a kd-tree method with
32 cells in a three-dimensional color-color-magnitude space.

The resulting catalog contains $1.2\times 10^6$ galaxies with
photometric redshifts. This catalog was matched against the catalog of
galaxies for which shapes were measured. This resulted in a sample of
$8\times 10^5$ galaxies with $18<R_C<24$, that are used in the
analysis presented here.

Comparison with the spectroscopic redshifts shows that accurate
photometric redshifts, with $\sigma_z<0.06$, can be derived in the
range $0.2<z<0.5$. At lower redshifts, the lack of $U$ band data
limits the accuracy, whereas at higher redshifts photometric errors
increase the scatter to $\sigma_z\sim 0.12$ (see Hsieh et al. 2005 for
more details).

\begin{figure}
\begin{center}
\leavevmode
\hbox{%
\epsfxsize=8cm
\epsffile[15 160 575 700]{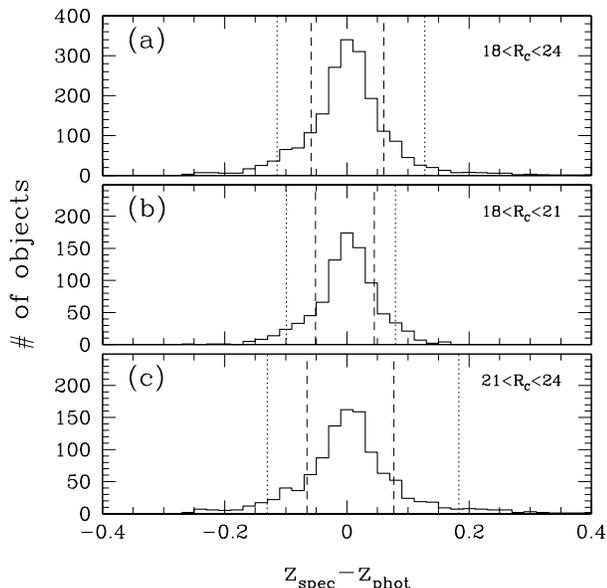}}
\caption{\footnotesize {\it panel a}: The difference in
spectroscopic and photometric redshifts for galaxies in the training
set, with $18<R<24$ and photometric redshifts $0.2<z_{\rm
phot}<0.4$. Our sample of lenses is selected to be in this redshift
and magnitude range. The dotted lines indicate the intervals
containing 90\% of the galaxies and the dashed lines indicate the 70\%
interval. {\it panel b}: Same, but now for the brighter half of the
training set, i.e., galaxies with $18<R<21$. {\it panel c}: Same, but
for the galaxies with $21<R<24$, the fainter half of the lenses.
\label{dzlens}}
\end{center}
\end{figure}

To study the halos of galaxies as a function of color and luminosity
we select a sample of lenses at intermediate redshifts: we select
galaxies with photometric redshifts $0.2<z<0.4$ and $R_C$-band
magnitudes $18<R_C<24$. This redshift range is well covered by the
CNOC2 redshift survey at the bright end, and the redshift errors 
are relatively small.  For the background galaxies we limit the
analysis to galaxies with $z_{\rm phot}<1$.

Figure~\ref{dzlens} shows the difference between spectroscopic and
photometric redshifts for different subsets of galaxies with
photometric redshifts $0.2<z<0.4$. Panel~a shows the full sample,
whereas panels b and c show the bright and faint halves respectively.
The distribution is peaked, with 70\% of the galaxies within the range
$|\Delta z|<0.06$ (0.05 and 0.07 for the bright and faint subsets,
resp.) and 90\% within $|\Delta z|<0.12$ (0.085 and 0.15 for the
bright and faint subsets, resp.). 

The solid histogram in Figure~\ref{zdist}a shows the normalized
photometric redshift distribution for the galaxies brighter than
$R_C=24$. It is common to parametrize the redshift distribution, and a
useful form is given by

\begin{equation}
p(z)=\frac{\beta}{z_s\Gamma[(1+\alpha)/\beta]}\left(\frac{z}{z_s}\right)^\alpha
\exp\left[-\left(\frac{z}{z_s}\right)^\beta\right].
\end{equation}

We fit this model to the observed redshift distribution. However, the
uncertainties in the photometric redshift determinations can be
substantial, and as a result the observed distribution is
broadened. We use the observed error distribution, assuming a normal
distribution, to account for the redshift errors. For the best fit
parameterization we find values of $z_s=0.29$, $\alpha=2$ (fixed) and
$\beta=1.295$, which yields a mean redshift of $\langle
z\rangle=0.53$. This model redshift distribution (which includes the
smoothing by redshift errors) is indicated by the smooth curve in
Figure~\ref{zdist}a.

In the weak lensing analysis, objects are weighted by the inverse
square of the uncertainty in the shear measurement (e.g., see Hoekstra
et al. 2000, 2002a). As more distant galaxies are fainter, they tend
to have somewhat lower weights and the effective redshift distribution
is changed slightly. The dashed histogram in Figure~\ref{zdist}a shows
the distribution weighted by the uncertainty in the shape measurement
for each redshift bin. The best fit parameterized redshift
distribution has parameters $z_s=0.265$, $\alpha=2.2$ (fixed) and
$\beta=1.30$, which yields $\langle z\rangle=0.51$, only slightly
lower than the unweighted case.

\subsection{Implications for cosmic shear results?}

Hoekstra et al. (2002b) presented constraints on the matter density
$\Omega_m$ and the normalization of the power spectrum $\sigma_8$ by
comparing cold dark matter predictions to the observed lensing signal
caused by large scale structure. The derived value for $\sigma_8$
depends critically on the adopted redshift distribution. Hoekstra et
al.  (2002b) used galaxies with $22<R_C<24$ and a redshift
distribution given by $z_s=0.302$, $\alpha=4.7$ and $\beta=1.7$, which
yields a mean redshift of $\langle z\rangle=0.59$. These parameters
were based on a comparison with redshift distributions determined from
the Hubble Deep Fields.

\begin{figure*}[!t]
\begin{center}
\leavevmode
\hbox{%
\epsfxsize=8.6cm
\epsffile[15 160 575 700]{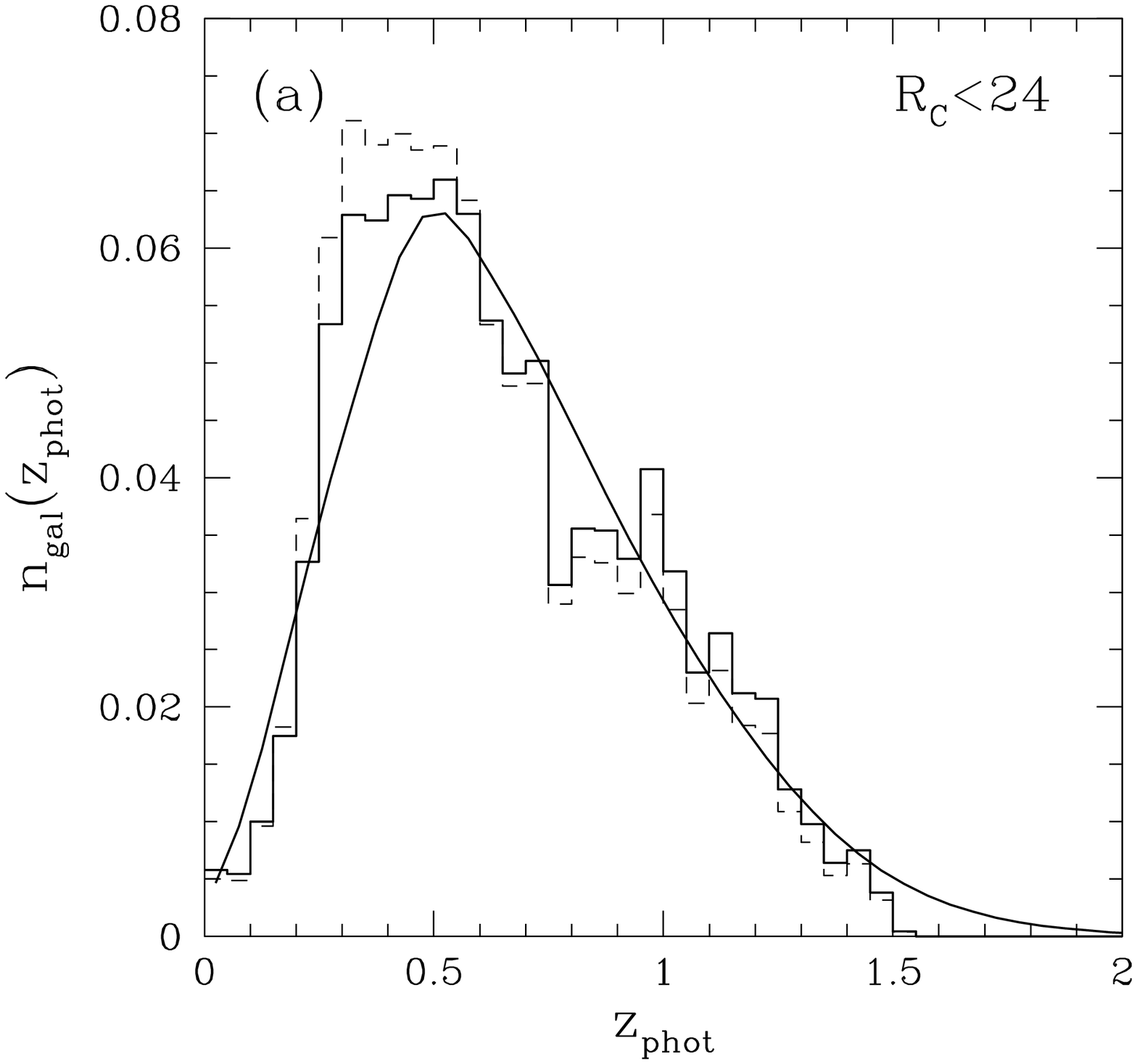}
\epsfxsize=8.6cm
\epsffile[15 160 575 700]{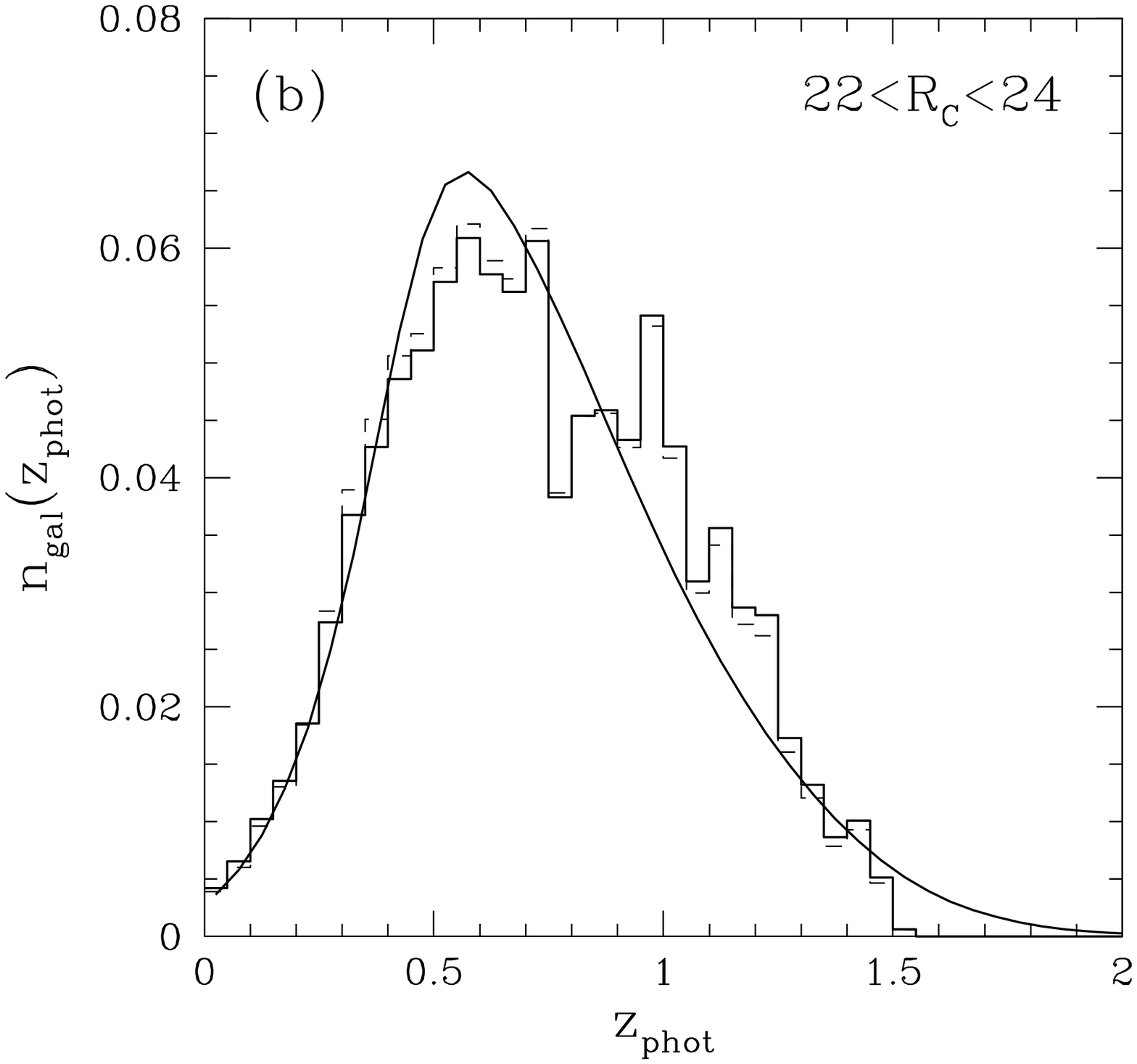}}
\caption{\footnotesize {\it panel a:} The solid histogram shows the
normalized photometric redshift distribution for the galaxies with
redshifts and magnitudes $R_C<24$ that are included in the weak
lensing analysis. The solid smooth curve shows the best fit model
redshift distribution (see text for details). {\it panel b:} Similar
to panel~a, but for galaxies with $22<R_C<24$, corresponding to the
range used by Hoekstra et al. (2002b). It is important to note that
the lack of a relatively good training set for $z>0.6$ limits the
interpretation.  The dashed histogram shows the distributions weighted
by the uncertainty in the shape measurement for each redshift bin.
\label{zdist}}
\end{center}
\end{figure*}

It is useful to examine how these assumptions compare to the RCS
photometric redshift distribution for galaxies with $22<R_C<24$, as
displayed in Figure~\ref{zdist}b. The best fit model, indicated by the
smooth curve, has parameters $z_s=0.31$, $\alpha=3.50$ and
$\beta=1.45$, implying a mean redshift of 0.65, about 10\% higher than
used by Hoekstra et al. (2002b). It is important to note, however,
that the training set lacks a large number of objects beyond $z=0.8$
and $R_C>22$. Despite these shortcomings, the mean redshift of sources
appears higher than what was used in Hoekstra et al. (2002b), thus
suggesting that their value for $\sigma_8$ needs to be revised
downwards. The suggested change in source redshift could reduce the
value for $\sigma_8$ from Hoekstra et al. (2002b) by about $8\%$ to
$\sigma_8\sim 0.8$. Unfortunately it is not possible to robustly
quantify the size of the revision. We stress that without further work
on photometric redshifts for faint, high redshift galaxies, it will be
difficult to interpret current and, most importantly, future cosmic
shear results.

The galaxy-galaxy lensing signal examined in this paper is much less
sensitive to the uncertainty in the redshift distribution of faint,
distant galaxies, as most of the signal is caused by lenses at much
lower redshifts. As mentioned above, to minimize uncertainties in our
results further, we only use background galaxies with redshifts less
than 1, and select a sample of lenses with redshift $0.2<z<0.4$.

\section{Testing the photometric redshifts}

Hsieh et al. (2005) present various tests of the accuracy of the
photometric redshifts. Comparison to the available spectroscopic data
as well as comparing to other published distributions provides a clear
way to quantify the uncertainties. In this section we discuss some
additional tests, based on the fact that the amplitude of the lensing
signal is a well known function of the source redshift. Such a test
provides a useful ``sanity'' check on the validity of the photometric
redshift distribution.

The azimuthally averaged tangential shear $\langle\gamma_t\rangle$
as a function of distance from the lens is a useful measure of the
lensing signal (e.g., Miralda-Escud{\'e} 1991):

\begin{equation}
\langle\gamma_t\rangle(r)=\frac{\bar\Sigma(<r) - 
\bar\Sigma(r)}{\Sigma_{\rm crit}}=\bar\kappa(<r)-\bar\kappa(r),
\end{equation}
 
\noindent where $\bar\Sigma(<r)$ is the mean surface density within an
aperture of radius $r$, and $\bar\Sigma(r)$ is the mean surface
density on a circle of radius $r$. The convergence $\kappa$, or
dimensionless surface density, is the ratio of the surface density and
the critical surface density $\Sigma_{\rm crit}$, which is given by

\begin{equation}
\Sigma_{\rm crit}=\frac{c^2}{4\pi G}\frac{D_s}{D_l D_{ls}},
\end{equation}
 
\noindent where $D_l$ is the angular diameter to the lens. $D_{s}$ and
$D_{ls}$ are the angular diameter distances from the observer to the
source and from the lens to the source, respectively. It is convenient
to define the parameter

\begin{equation}
\beta=\max[0,D_{ls}/D_s],
\end{equation}

\noindent which is a measure of how the amplitude of the lensing signal
depends on the redshifts of the source galaxies. For instance, in the
case of a singular isothermal sphere (SIS) model, the dimensionless
surface density is

\begin{equation}
\kappa=\gamma_t=\frac{r_E}{2r},
\end{equation}

\noindent where $r_E$ is the Einstein radius. Under the assumption of
isotropic orbits and spherical symmetry, the Einstein radius (in
radians) is related to the velocity dispersion and $\beta$ through

\begin{equation}
r_E=4\pi\left(\frac{\sigma}{c}\right)^2 \beta.
\end{equation}

To test the photometric redshifts from the RCS, we use galaxies with
photometric redshifts $0.2<z<0.4$ to define a sample of lenses. We
compute the ensemble averaged tangential shear around these galaxies
(i.e., the galaxy-mass cross-correlation function) as a function of
source redshift. Brighter galaxies are expected to be more massive,
and should be given more weight. To derive the lensing signal, we
assume that the velocity dispersion scales with luminosity as
$\sigma\propto L_B^{0.3}$, a choice which is motivated by the oberved
slope of the $B$-band Tully-Fisher relation (e.g., Verheijen 2001).

\begin{figure}
\begin{center}
\leavevmode
\hbox{%
\epsfxsize=8cm
\epsffile[15 160 575 700]{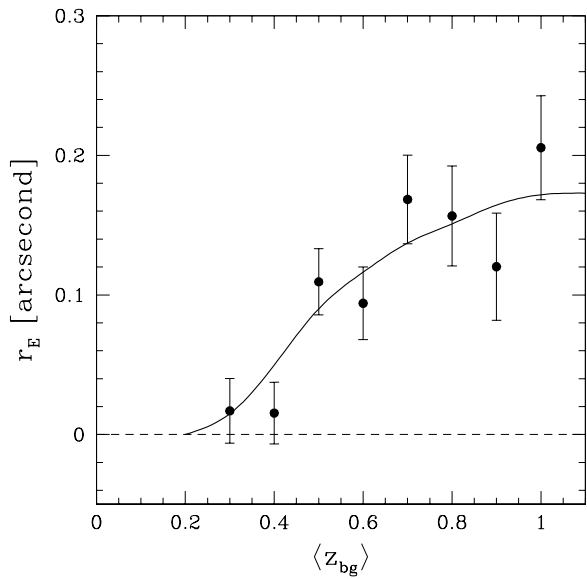}}
\caption{\footnotesize Best fit Einstein radius obtained from a fit
to the tangential shear as a function of redshift of the background
galaxies. Lenses were selected to have photometric redshifts in the
range $0.2<z<0.4$. The solid line corresponds to the dependence of
the lensing signal for a $\Lambda$CDM cosmology. The observed lensing
signal scales with redshift as expected.
\label{re_zbg}}
\end{center}
\end{figure}

We select bins with a width of 0.1 in redshift, and measure the
galaxy-mass cross-correlation function (e.g., see Hoekstra et
al. 2004; Sheldon et al. 2004) out to 10 arcminutes. This signal
arises from the combination of the clustering properties of the lenses
and the underlying dark matter distribution. In the remainder of the
paper, while studying the properties of dark matter halos around
galaxies, we limit the analysis to smaller radii and to `isolated'
lenses. However, by extending the range of measurements in this
section, the signal-to-noise ratio is higher. The signal is well
described by a SIS model for this range of scales (as suggested by the
reduced $\chi^2$ values for the fits). The resulting value for the
Einstein radius as a function of redshift for the background galaxies
is presented in Figure~\ref{re_zbg}.

We find a negligible lensing signal for galaxies at the redshift of
the lenses, whereas it increases for more distant sources. For a given
cosmology and a pair of lens and source redshifts the value of $\beta$
can be readily computed. However, the errors in the photometric
redshift determination complicate such a simple comparison between the
expected signal and the results presented in Figure~\ref{re_zbg}.  As
was the case for the photometric redshift distribution, the redshift
errors will change the signal. For instance at low redshifts, higher
redshift galaxies will scatter into this bin, thus increasing the
lensing signal. At higher redshifts, lower redshift object will
scatter upwards, lowering the signal.

When comparing the observed signal to the signal expected based on the
adopted $\Lambda$CDM cosmology we need account for these redshift
errors. To this end, we create simulated catalogs. The first step is
to compute a model lensing signal based on the observed photometric
redshifts (which are taken to be exact). We then create a mock catalog
by adding the random errors to the redshift (while leaving the lensing
signal unchanged). These random errors are based on the observed
distribution (see e.g., Fig.~\ref{dzlens}). We measure the lensing
signal as a function of redshift in the mock catalog. This signal,
indicated by the solid line in Figure~\ref{re_zbg} can be compared
directly to our actual measurements as it now includes the smoothing
effect of redshift errors. Figure~\ref{re_zbg} shows that it traces
the observed change in amplitude of the lensing signal very well.

Another useful experiment is to measure the lensing signal when the
lenses and sources are in the same redshift bin. We note that this
procedure enhances the probability that we measure the signal for
galaxies which are physically associated. If satellite galaxies tend
to be aligned tangentially (radially) this would also lead to a
positive (negative) signal. The results from Bernstein \& Norberg
(2002), based on an analysis employing spectroscopic redshifts, has
shown that intrinsic tangential alignments are negligible.
Unfortunately, in our case, the much larger photometric redshift
errors (compared to spectroscopic redshifts) effectively suppress this
potentially interesting signal. In addition, for the RCS data, the
interpretation of the signal requires a large set of spectroscopic
redshifts to quantify the contributions of unassociated galaxies in
each bin.

\begin{figure}
\begin{center}
\leavevmode
\hbox{%
\epsfxsize=8cm
\epsffile[15 160 575 700]{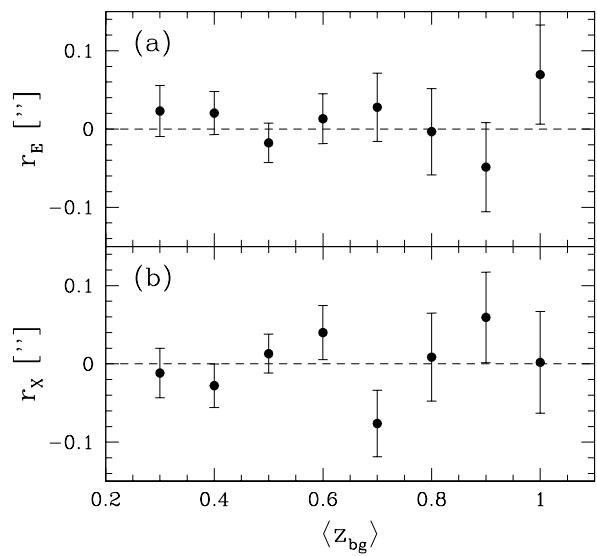}}
\caption{\footnotesize {\it panel a}: Best fit Einstein radius
obtained from a fit to the tangential shear when the lenses and
sources are selected to be in the same redshift bin. In this case, no
signal should be present, in agreement with the measurements. This
result also indicates that intrinsic tangential alignments are
neglibible. {\it panel b}: Results when the background galaxies
are rotated by 45 degrees (`B'-mode). Also in this case no signal
is detected. \label{re_zbin}}
\end{center}
\end{figure}

Instead, we use these measurements as a test of the photometric
redshifts.  The results are presented in Figure~\ref{re_zbin}. Panel~a
shows the results for the tangential shear, whereas Panel~b shows the
results when the background galaxies are rotated by 45$^\circ$ (which
is a measure of systematics). In both cases, we do not observe a
significant signal. The lack of a signal in the tangential shear in
this case implies that the errors in photometric redshifts are
relatively small. If this were not the case, and higher redshift
galaxies would contaminate the samples at lower redshifts, we would
expect to observe a positive signal.

\section{Galaxy dark matter profile}

One of the major advantages of weak gravitational lensing over
dynamical methods is that the lensing signal can be measured out to
large projected distances from the lens. However, at large radii, the
contribution from a particular galaxy may be small compared to its
surroundings: a simple interpretation of the measurements can only be
made for `isolated' galaxies. 

In practice, galaxies are not isolated, which is particularly true for
bright, early-type galaxies. In their analysis of SDSS data, Guzik \&
Seljak (2002) quantified the contribution from clustered galaxies
using a halo-model approach. As discussed in \S5, we follow a
different approach by selecting relatively isolated galaxies. As a
result, our results are not strictly valid for the galaxy population
as a whole. Nevertheless, the selection procedure is well defined and
can be readily implemented when comparing to numerical simulations.

We limit the analysis to relatively small distances from the lens,
thus ensuring that the signal is dominated by the lens itself. As a
result, we need to adopt a model for the mass distribution to relate
the lensing signal to the mass of the lens. Our choice is motivated by
the results of cold dark matter (CDM) simulations.

Collisionless cold dark matter provides a good description for the
observed structures in the universe. Numerical simulations, which
provide a powerful way to study the formation of structure in the
universe, indicate that on large scales CDM gives rise to a particular
density profile (e.g., Dubinski \& Carlberg 1991; Navarro, Frenk, \&
White 1995, 1996, 1997; Moore et al. 1999). We note, however, that
there are still uncertainties regarding the slope at small radii and
the best analytical description of the profile (e.g., Moore et
al. 1999; Diemand et al. 2004; Hayashi et al. 2004; Tasitsiomi et
al. 2004a).  Furthermore there is considerable scatter from halo to
halo in the simulations. Our observations, however, cannot distinguish
between these various profiles, and instead we focus on the commonly
used NFW profile, given by

\begin{equation}
\rho(r)=\frac{M_{\rm vir}}{4\pi f(c)}\frac{1}{r(r+r_s)^2},
\end{equation}

\noindent where $M_{\rm vir}$ is the virial mass, which is the mass
enclosed within the virial radius $r_{\rm vir}$. The virial radius is
related to the `scale radius' $r_s$ through the concentration $c=r_{\rm
vir}/r_s$. The function $f(c)=\ln(1+c)-c/(1+c)$.
 
One can fit the NFW profile to the measurements with $M_{\rm vir}$ and
concentration $c$ (or equivalently $r_s$) as free parameters. However,
numerical simulations have shown that the average concentration
depends on the halo mass and the redshift. Hoekstra et al. (2004)
constrained the mass and scale radius of the NFW model using a maximum
likelihood analysis of the galaxy-galaxy lensing signal, and found
that the results agreed well with the predictions from simulations.
We therefore adopt the results from Bullock et al. (2001), who found
from simulations that
 
\begin{equation}
c=\frac{9}{1+z}\left(\frac{M_{\rm vir}}{8.12\times 10^{12} h
M_\odot}\right)^{-0.14}.
\end{equation}

\noindent It is good to note that individual halos in the simulations
have a lognormal dispersion of approximately 0.14 around the median.
For the virial mass estimates presented here, we will use this
relation between mass and concentration, thus assuming we can describe
the galaxy mass distribution by a single parameter. 

\noindent By definition, the virial mass and radius are related by
 
\begin{equation}
M_{\rm vir}=\frac{4\pi}{3} \Delta_{\rm vir}(z)\rho_{\rm bg}(z)r_{\rm vir}^3,
\end{equation}
 
\noindent where $\rho_{\rm bg}=3H_0^2\Omega_m(1+z)^3/(8\pi G)$ is the
mean density at the cluster redshift and the virial overdensity
$\Delta_{\rm vir}\approx (18\pi^2+82\xi-39\xi^2)/\Omega(z)$, with
$\xi=\Omega(z)-1$ (Bryan \& Norman 1998). For the $\Lambda$CDM
cosmology considered here, $\Delta_{\rm vir}(0)=337$. We also note
that for the adopted $\Lambda$CDM cosmology the virial mass is
different from the widely used $M_{200}$. This mass is commonly
defined as the mass contained within the radius $r_{200}$, where the
mean mass density of the halo is equal to $200\rho_c$ (i.e., setting
$\Delta=200$ and $\rho_{\rm bg}=\rho_c$ in Eqn.~9). Note, however,
that other definitions for $M_{200}$ can be found in the literature as 
well.

The expressions for the tangential shear and surface density for the
NFW profile have been derived by Bartelmann (1996) and Wright \&
Brainerd (2000) and we refer the interested reader to these papers for
the relevant equations.

\section{Results}

As discussed above, we study a sample of lenses with photometric
redshifts $0.2<z<0.4$ and $18<R_C<24$. This first selection yields
$\sim 1.4\times 10^5$ lenses. We split this sample in a number of
luminosity and color bins and determine the virial radii from an NFW
model fit to the observed lensing signal.

For bright galaxies the lensing signal on small scales is typically
dominated by the dark matter halo associated with that galaxy. In the
case of faint (low mass) galaxies, however, the signal can easily be
dominated by contributions from a massive neighbor. Note that this
neighbor need not be physically associated with the lens, as all
matter along the line-of-sight contributes to the lensing signal.

We can study the relevance of the local (projected) density by
measuring the lensing signal around a sample of `faint' lenses
($10^9<L_B<5\times 10^9~h^{-2} L_{B\odot}$), as a function of the
projected distance to the nearest `bright' lens ($L_B>5\times 10^9
h^{-2}L_{B\odot}$). This distance can be used as a crude measure of
the density around the faint lens (i.e., the smaller the distance, the
higher the density).

\begin{figure}
\begin{center}
\leavevmode
\hbox{%
\epsfxsize=8cm
\epsffile[15 160 575 700]{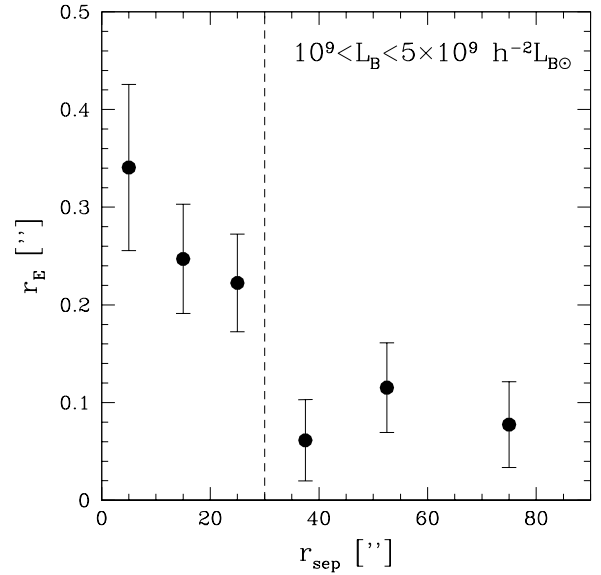}}
\caption{\footnotesize Einstein radius for `faint' lenses as a
function of projected distance to the nearest `bright' lens $r_{\rm
sep}$. The faint galaxies have luminosities $10^{9}<L_B<5\times 10^9
h^{-2}\lbsun$, whereas the bright galaxies have $L_B>5\times 10^9h^{-2}
\lbsun$.\label{re_dist}}
\end{center}
\end{figure}

To this end, we split this sample of `faint' lenses into subsets based
on their distance to the nearest bright galaxy. We fit a SIS model to
the ensemble averaged lensing signal out to 2' ($\sim 400h^{-1}$kpc at
the mean distance of the lenses) for each bin. The reduced $\chi^2$
values for the best fit are all close to unity, indicating that the
SIS model provides a good fit to these observations. We found that
limiting the fit to smaller radii did not change the results apart
from increasing the measurement errors. Figure~\ref{re_dist} shows the
derived value for the Einstein radius as a function of the distance to
the nearest bright galaxy.

\begin{figure*}[!th]
\begin{center}
\leavevmode 
\hbox{%
\epsfxsize=\hsize 
\epsffile[20 170 560 460]{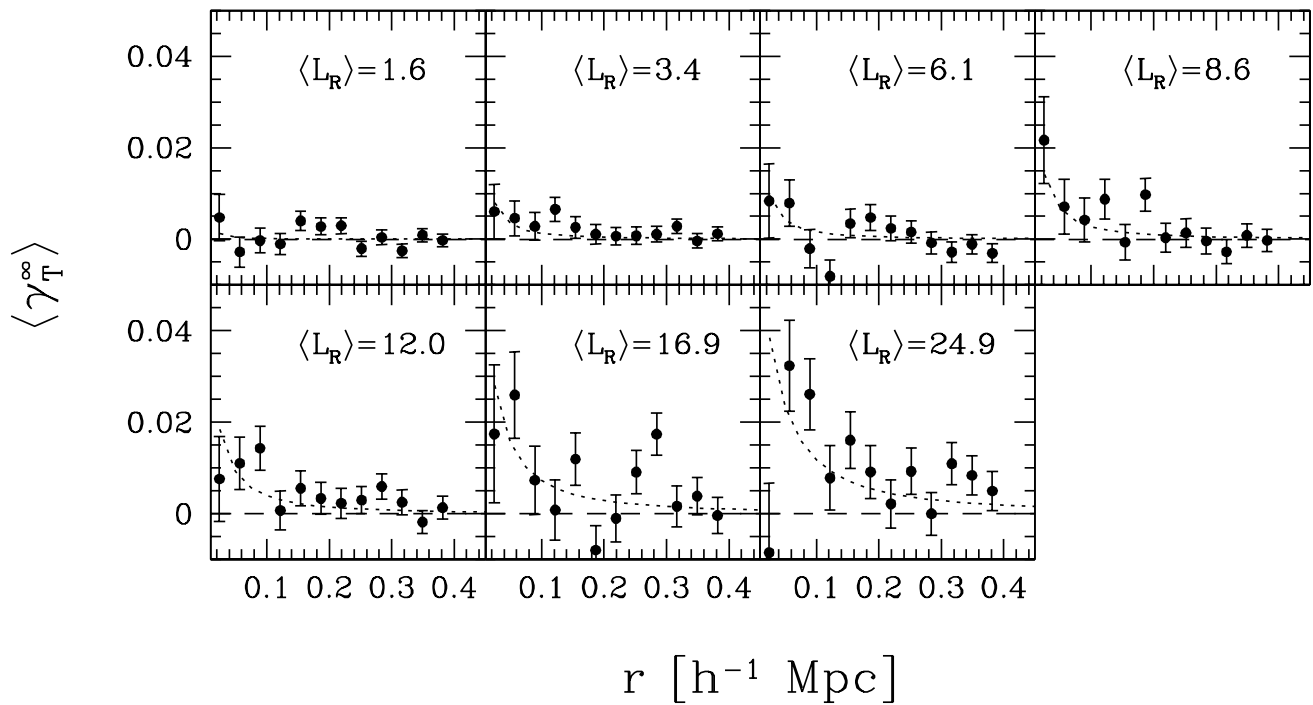}} 
\caption{\footnotesize Tangential shear as a function of projected
(physical) distance from the lens for each of the seven restframe
$R$-band luminosity bins. To account for the fact that the lenses have
a range in redshifts, the signal is scaled such that it corresponds to
that of a lens at the average lens redshift ($z\sim 0.32$) and a
source redshift of infinity The mean restframe $R$-band luminosity for
each bin is also shown in the figure in units of $10^9
h^{-2}$L$_{R\odot}$.  The strength of the lensing signal clearly
increases with increasing luminosity of the lens. The dotted line
indicates the best fit NFW model to the data. The tangential shear
profiles for the $B$ and $V$-band are very similar and we only present
the final results for the $R$ filter in Figure~\ref{ml_all}
\label{gtprof}}
\end{center}
\end{figure*}

The results show a clear increase in lensing signal as the separation
decreases, i.e. as the density increases. However, at separations
larger than $\sim 30''$, the observed lensing signal appears to be
independent of the density. Larger data sets are required to make more
definitive statements, but these findings suggest that we can measure
the properties of `isolated' faint galaxies by limiting the sample to
galaxies which are more than 30 arcseconds away from a brighter
galaxy. Although this is a rather strict selection for the faintest
galaxies, bright galaxies can be surrounded by many faint galaxies and
consequently are not truely isolated. In the remainder of this paper,
we present results based on the sample of `isolated' galaxies, unless
specified otherwise. This selection reduces the sample of lenses to
94,509 galaxies.

\subsection{Mass-luminosity relation}

We split the sample of `isolated' lens galaxies into seven luminosity
bins and measure the mean tangential distortion as a function of
radius out to 2 arcminutes. We fit an NFW profile to these
measurements, with the virial mass as a free parameter, as described
in \S4. 

Figure~\ref{gtprof} shows the measurements of the tangential shear as
a function of projected distance from the lens for the seven $R$-band
luminosity bins. The results for the $B$ and $V$ filters are very
similar to the ones presented in Figure~\ref{gtprof}. To account for
the fact that the lenses span a range in redshift, we have scaled the
signal such that it corresponds to that of a lens at the mean lens
redshift ($z\sim 0.32$) and a background galaxy at infinite
redshift. In each panel in Figure~\ref{gtprof} the average restframe
$R$-band luminosity for each bin is indicated (in units of
$10^{9}h^{-2}$L$_{R\odot}$). The vertical scales in each of the panels
in Figure~\ref{gtprof} are the same, and as the luminosity of the
lenses increases we observe a clear increase in the strength of the
lensing signal. The best fit NFW models for each bin are indicated by
the dotted curves. Note that the current observations cannot
distinguish between an NFW profile (used here) and other profiles
such as the SIS model.

There are more faint galaxies relative to the number of bright
galaxies and the errors in the photometric redshift estimates will
have as the net effect of faint galaxies getting scattered to higher
luminosity bins, hence biasing the mass at fixed luminosity to a lower
value. To estimate the level of this bias we create mock catalogs. We
assume a power-law mass-luminosity relation and compute the model
lensing signal using the observed photometric redshifts of the lens
and source galaxies. We analyse this `perfect' catalog and compute the
virial masses as a function of luminosity. We then use the observed
photometric redshift error distribution as a function of apparent
magnitude (see Figure~\ref{dzlens}) to create a number of new catalogs
where the random error is added to the redshift (note that the lensing
signal is not changed). These catalogs are also analysed and yield the
`observed' virial mass as a function of luminosity.

\begin{figure}
\begin{center}
\leavevmode
\hbox{%
\epsfxsize=8.5cm
\epsffile{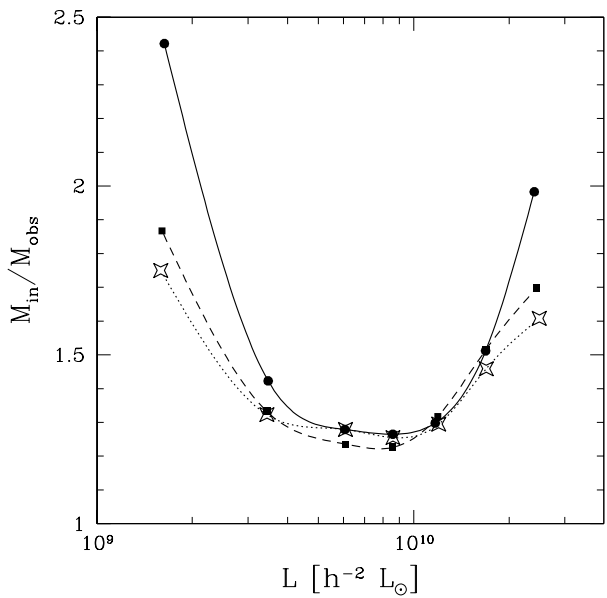}}
\caption{\footnotesize The ratio of the input virial mass and the
observed mass after adding photometric redshift errors. The dependence
with luminosity is dominated by how the redshift errors depend on
brightness. The resulting curves depend only very weakly on the input
mass-luminosity relation. The corrections are somewhat different for the
various restframe bands. The solid line with filled circles correspond
to the $B$-band results, the dashed line with solid squares is for the
$V$-band and the dotted line with stars is for the $R$-band data. To
infer the correct virial mass, we scale the observed virial masses by
these curves. \label{corfac}}
\end{center}
\end{figure}

As expected, the resulting masses are smaller than the input masses
and the change in mass depends on the luminosity. The results are
presented in Figure~\ref{corfac} for the $B$, $V$, and
$R$-band. Different choices for the mass-luminosity relation (within
reasonable bounds) yield very similar curves. To infer the correct
virial mass, we scale the observed virial masses by these curves.

\begin{table*}
\begin{center}
\caption{Best fit virial masses\label{tab_vir}}
\begin{tabular}{cc|cc|cc}
\hline
\hline
$L_B$ & $M_{\rm vir}$ & $L_V$ & $M_{\rm vir}$ & $L_R$ & $M_{\rm vir}$\\ 
$[10^9 h^{-2}{\rm L}_\odot] $ & $[10^{11}h^{-1}{\rm M}_\odot]$ & 
$[10^9 h^{-2}{\rm L}_\odot] $ & $[10^{11}h^{-1}{\rm M}_\odot]$ & 
$[10^9 h^{-2}{\rm L}_\odot] $ & $[10^{11}h^{-1}{\rm M}_\odot]$ \\
\hline
1.6   & $0.66^{+0.41}_{-0.43}$ &  1.6  & $0.48^{+0.33}_{-0.35}$ &  1.6  & $0.10^{+0.40}_{-0.30}$ \\
3.5   & $0.86^{+0.42}_{-0.48}$ &  3.4  & $1.05^{+0.69}_{-0.45}$ &  3.4  & $1.24^{+0.65}_{-0.57}$ \\
6.1   & $1.81^{+0.84}_{-0.75}$ &  6.1  & $3.1^{+1.2}_{-1.0}$    &  6.1  & $1.62^{+0.88}_{-0.84}$ \\ 
8.6   & $6.0^{+1.6}_{-1.6}$    &  8.6  & $2.6^{+1.3}_{-1.1}$    &  8.6  & $3.1^{+1.4}_{-1.3}$    \\
11.7  & $7.7^{+2.0}_{-1.9}$    &  11.9 & $6.6^{+1.9}_{-1.8}$    &  12.0 & $5.0^{+1.9}_{-1.5}$    \\
16.9  & $16.9^{+5.5}_{-4.9}$   &  17.0 & $20.1^{+5.3}_{-4.9}$   &  16.9 & $11.5^{+4.0}_{-3.4}$   \\
24.0  & $18.8^{+7.8}_{-6.3}$   &  24.5 & $17.2^{+5.6}_{-4.9}$   &  24.9 & $23.3^{+5.6}_{-5.1}$   \\
\hline
\end{tabular}
\end{center}
\tablecomments{Best fit virial masses as a function of luminosity in
the restframe $B$, $V$ and $R$ band. The corresponding values for the
concentration $c$ can be computed using Eqn.~8 using a redshift of
$z=0.32$ for the lenses.  The listed errors indicate the 68\%
confidence limits.}
\end{table*}

At the low luminosity end the corrections are large because of the
relatively large errors in redshift. At the bright end, however, the
redshift errors are smaller, but the number of bright galaxies is
decreasing rapidly (because of the shape of the luminosity function),
and a relatively larger fraction of intrinsically lower mass systems
ends up in the high luminosity bin, resulting in an increase of the
correction factor. The corrections are substantial at both ends, but
the origin is well understood and the associated uncertainty is small.

\begin{figure*}[!t]
\begin{center}
\leavevmode 
\hbox{%
\epsfxsize=\hsize 
\epsffile[50 334 570 685]{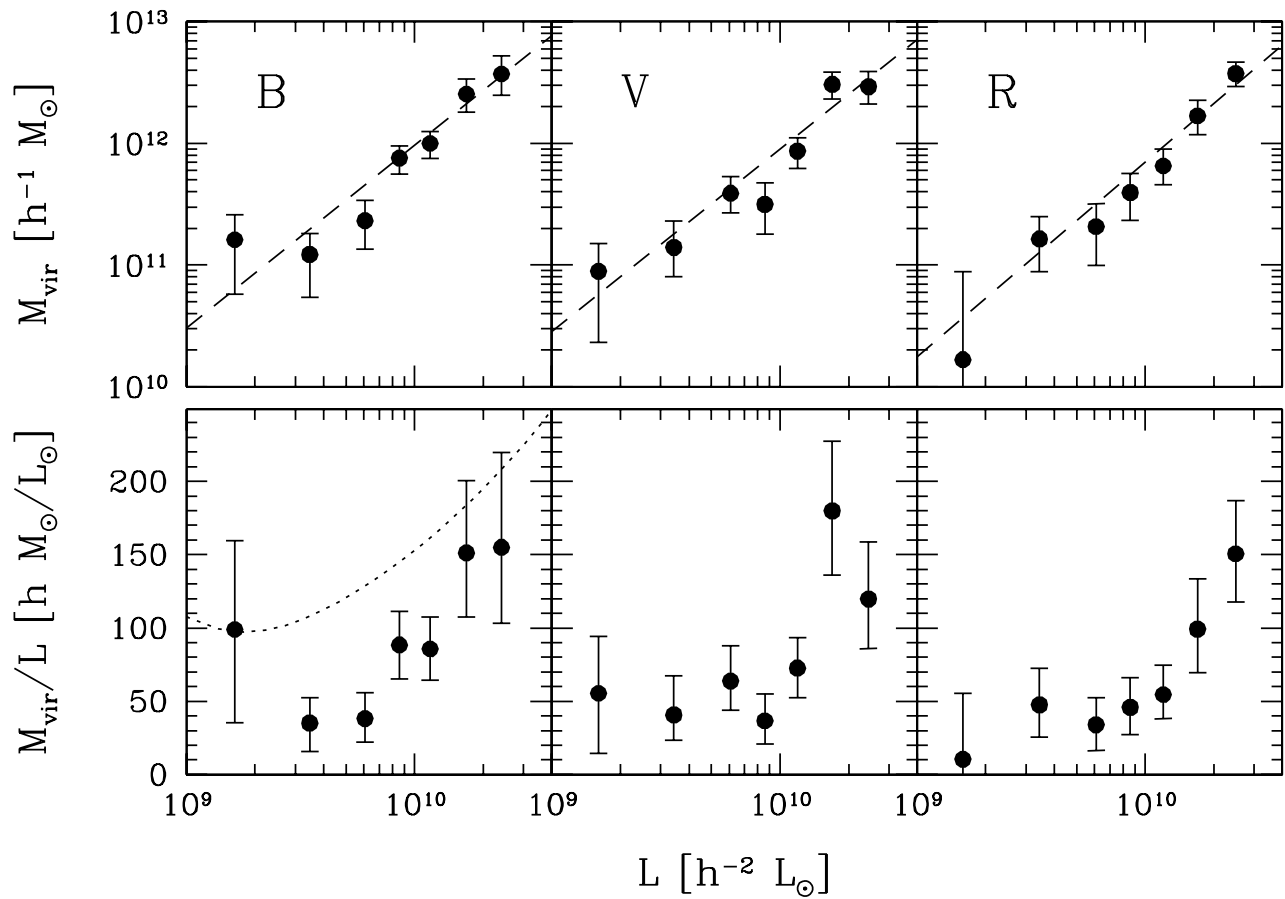}} 
\caption{\footnotesize {\it upper panels}: Virial mass as a function
of the rest-frame luminosity in the indicated filter. The dashed line
indicates the best fit power-law model for the mass-luminosity
relation, with the relevant parameters listed in
Table~\ref{tab_mass}. {\it lower panels:} Observed rest-frame virial
mass-to-light ratios. The results suggest a rise in the mass-to-light
ratio with increasing luminosity, albeit with low significance. The
dotted line in the panel showing the $B$-band mass-to-light ratio
corresponds to model~A from van den Bosch et al. (2003). It matches
the observed dependence of the mass-to-light ratio with luminosity,
but with an offset towards higher values.
\label{ml_all}}
\end{center}
\end{figure*}

The corrected virial masses as a function of luminosity in the $B$,
$V$, and $R$-band respectively, are presented in the upper panels of
Figure~\ref{ml_all} and the best fit virial masses are listed in
Table~\ref{tab_vir}. In all cases we see a clear increase of the
virial mass with luminosity. The results suggest a power-law relation
between the luminosity and the virial mass, although this assumption
might not hold at the low luminosity end (e.g., see lower panels of
Fig.~\ref{ml_all}). We therefore fit

\begin{equation}
M=M_{\rm fid}\left(\frac{L}{10^{10} h^{-2} \lsun}\right)^\alpha,
\end{equation}

\noindent to the measurements, where $M_{\rm fid}$ is the virial mass
of a fiducial galaxy of luminosity $L=10^{10}h^{-2}$L$_{x\odot}$, where
$x$ indicates the relevant filter. The best fit in each
filter is indicated by the dashed line in Figure~\ref{ml_all}. The
resulting best fit parameters for this mass-luminosity relation are
listed in Table~\ref{tab_mass}. We do not observe a change in the
slope $\alpha$ for the different filter, but find that $M_{\rm fid}$
is decreasing for redder passbands.

Tasitsiomi et al. (2004b) studied the weak lensing mass-luminosity
relation from their numerical simulations. This study shows that the
interpretation of the mass-luminosity relation presented in
Figure~\ref{ml_all} is complicated by the fact that the halos of
galaxies of a given luminosity show a scatter in their virial masses.
For the model adopted in Tasitsiomi et al. (2004b), the best fit
virial mass gives a value between the median and mean mass. The
amplitude of this bias depends on the assumed intrinsic scatter in the
mass-luminosity relation, which requires further study. The Tasitsiomi
et al. (2004b) results imply that our results underestimate the actual
mean virial mass, but that the slope of the mass-luminosity relation
is not changed.

Guzik \& Seljak (2002) measured the mass-luminosity relation using
data from the SDSS. The average luminosity of their sample of lenses
is higher than studied here.  Also, the analysis by Guzik \& Seljak
(2002) differs from ours, as they model the contribution from other
halos. Using the halo model approach they compute the contributions of
other halos to the lensing signal around a galaxy, including that of
smooth group/cluster halos. In this paper, we have instead minimized
such contributions to the galaxy-galaxy lensing signal by selecting
`isolated' galaxies and limiting the analysis to the lensing signal
within 400$h^{-1}$kpc from the lens. The results presented in
Figures~\ref{re_dist} and~\ref{gtprof} suggest that this approach has
worked well.

\begin{table}
\begin{center}
\caption{Best fit parameters of the mass-luminosity relation
\label{tab_mass}}
\begin{tabular}{lcc}
\hline
\hline
filter & $M_{\rm fid}$ & $\alpha$ \\
       & [$10^{11}h^{-1}$M$_\odot$] & \\
\hline
B     & $9.9^{+1.5}_{-1.3}$ & $1.5\pm0.3$ \\ 
V     & $9.3^{+1.4}_{-1.3}$ & $1.5\pm0.2$ \\
R     & $7.5^{+1.2}_{-1.1}$ & $1.6\pm0.2$ \\
\hline
\end{tabular}
\end{center}
\tablecomments{Column~2 lists the virial mass for a galaxy of luminosity
$10^{10}h^{-2}$L$_\odot$ in the indicated filter. Column~3 lists the best fit
power-law slope of the mass-luminosity relation. The listed errors
indicate the 68\% confidence limits.}
\end{table}

Guzik \& Seljak (2002) present results for two different cases of
group halo contributions. Depending on the assumed relative importance
of such a halo the derived mass changes only slightly. The assumptions
for the halo contribution do affect the inferred slopes somewhat,
although the change is small for the redder filters. Minimizing the
halo contribution yields a power-law slope of $\sim 1.5-1.7$, in
excellent agreement with the findings presented here. However, when
maximizing the effect the slope decreases to $\sim 1.3-1.4$ in the red
filters and to $1.2\pm0.2$ in the $g'$-band.  It should be noted that
the latter scenario is rather extreme, given that, with the exception
of the central galaxy, the relative importance of the group halo is
expected to diminish with increasing luminosity (i.e., mass) of the
lens. In addition, the different range in luminosities probed in the
two studies would most likely affect the results in the bluest
filters.

Benson et al. (2000) present predictions for the $B$-band
mass-luminosity relation based on semi-analytic models of galaxy
formation. In the luminosity range probed here, they obtain a
power-law slope of $\sim 1.6$. Van den Bosch et al. (2003) used the
conditional luminosity functions computed from the 2dF galaxy redshift
survey to constrain the variation of the mass-to-light ratio as a
function of mass. Van den Bosch consider a number of models, which
provide similar mass-luminosity relations for the range of masses
probed in this paper. We consider their model A, which is obtained by
fitting the data, without constraining the model parameters. For this
model the mass-luminosity relation is close to a power law with a
slope of 1.3. Hence, both model predictions are in good agreement with
our findings and the results of Guzik \& Seljak (2002).  

The agreement in the slope of the mass-luminosity relation strengthens
the conclusion by Guzik \& Seljak (2002) that rotation curves must
decline substantially from the optical to the virial radius, in order
to reconcile our results with the observed scaling relations at small
radii, such as the Tully-Fisher relation. A decrease in rotation
velocity is also predicted by semi-analytic models of galaxy formation
(e.g., Kauffmann et al. 1999; Benson et al. 2000).

Guzik \& Seljak (2002) define the virial mass in terms of an
overdensity of 200 times the critical density, which is different from
ours. They find a mass $M_{200}=(9.3\pm1.6)\times 10^{11}h^{-1}\msun$
for a galaxy with a luminosity of $1.1\times 10^{10}h^{-2}\mlg$ at a
redshift of $z\sim 0.16$.  We convert our mass estimate to their
definition, and use the transformations between filters from Fukugita
et al. (1996), to relate our results to those of Guzik \& Seljak
(2002). Furthermore, we assume that the fiducial galaxy is about 10\%
brighter at z=0.32, compared to $z=0.16$. Under these assumptions, our
results translate to a mass $M_{200}=(11.7\pm1.7)\times 10^{10}h^{-1}
\msun$ for a galaxy with a luminosity of $1.1\times
10^{10}h^{-2}\mlg$, in agreement with the findings of Guzik \& Seljak
(2002) at the $1\sigma$ level.

The lower panels in Figure~\ref{ml_all} show the inferred rest-frame
mass-to-light ratios as a function of luminosity for the different
filters. The results suggest a rise in mass-to-light ratio for
galaxies more luminous than $\sim 10^{10}h^{-2}$\lsun ~and little
variation for fainter galaxies.  This suggests that a power law is not
sufficient to describe the mass-luminosity relation over the range
probed here. However, a larger data set is needed to make a firm
statement. The dashed curve in Figure~\ref{ml_all} shows the values
corresponding to model~A from van den Bosch et al. (2003), converted
to the $B-$band and our definition of the virial mass.  The predicted
mass-to-light ratio is significantly higher compared to our
measurements. This could point to a systematic underestimate of the
virial masses from lensing due to scatter in the mass-luminosity
relation, as suggested by Tasitsiomi et al. (2004b). Nevertheless,
the model predictions are in qualitative agreement with the results
presented in Figure~\ref{ml_all}, in the sense that they predict a
rise for bright galaxies and  a small increase in mass-to-light
ratio towards lower luminosities.

\subsection{Star formation efficiency}

In the previous section we studied the dependence of the mass-to-light
ratio as a function of luminosity. The results suggest an increase for
luminous galaxies. A simple interpretation of these results, however,
is complicated because the mix of galaxy type is also a function of
luminosity. The more luminous galaxies are likely to be early-type
galaxies rather than spiral galaxies.

Although we have not classified our sample of lenses, we can use the
$B-V$ color as a fair indicator of galaxy type (e.g., Roberts \&
Haynes 1994). Furthermore, the color can be used to estimate the mean
stellar mass-to-light ratio, which also is a strong function of color
(e.g., Bell \& de Jong 2001). Comparison of the virial and stellar
mass-to-light ratios then enables us to estimate the relative fraction
of the mass that has been transformed into stars.

\begin{figure}
\begin{center}
\leavevmode
\hbox{%
\epsfxsize=8cm
\epsffile[15 160 575 700]{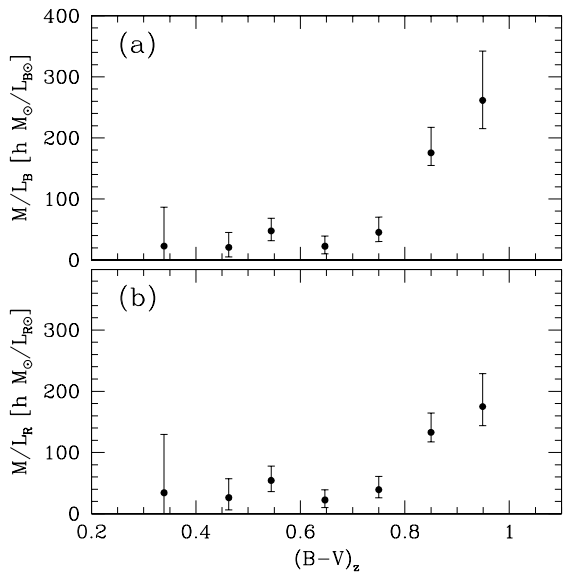}}
\caption{\footnotesize (a) Rest-frame $B$-band virial mass-to-light
ratio as a function of rest-frame $(B-V)$ color. (b) The same, but now
for the rest-frame $R$-band. In this case the change for red galaxies
is smaller.  The filled circles are the measurements for our sample of
lenses, which have a mean redshift $z=0.32$. 
\label{mlcol}}
\end{center}
\end{figure}

Figure~\ref{mlcol}a shows the inferred $B$-band mass-to-light ratio as
a function of restframe $B-V$ color. Figure~\ref{mlcol}a shows a clear
increase in mass-to-light ratios for early-type galaxies, which have
colors redder than $\sim 0.8$. It is useful to note that our selection
of lenses allows for the brightest galaxies in the centres of denser
regions to be included. Our simulations show that the inferred masses
are not biased, but these tests do not include the smooth
contributions from group halos. The resulting mass-to-light ratios are
comparable to those determined for rich clusters of galaxies (e.g.,
Hoekstra et al. 2002d) and massive galaxy groups (Parker et
al. 2005). For galaxies with $B-V<0.8$ the mass-to-light ratio does
not show a clear change with color and we find an average
mass-to-light ratio of $M/L_B=32\pm9\mlb$.  Figure~\ref{mlcol}b shows
the results for the $R$-band mass-to-light ratio. For galaxies with
$B-V<0.8$ we obtain an average value of $M/L_R=34\pm9\mlr$. The
increase in mass-to-light ratio for red galaxies is smaller in the
$R$-band, which is expected since the stellar mass-to-light ratios
also vary less.

As discussed in the previous section, the measurements presented in
Figure~\ref{mlcol} are for a sample of galaxies with an average
redshift of $z=0.32$. Note, that to compare these results to
measurements at lower redshifts one needs to account for evolution in
both the colors and the luminosities of the lens galaxies. To this end
we use population synthesis models (e.g., Fioc \& Rocca-Volmerange
1997) which indicate that the galaxies become redder as they age, and
that the reddest galaxies dim somewhat faster than the blue galaxies.

As mentioned above, it is interesting to estimate the fraction
of mass in stars. To do so, we need to relate the luminosity
to the stellar mass. Direct measurements of the stellar mass-to-light
ratios are difficult, although rotation curves can provide useful
limits on the maximum allowed value. Instead we rely on galaxy
evolution models, which use evolutionary tracks and assumptions
about initial mass function (IMF), the star formation history and
feedback, to compute stellar populations as a function of age.

There are many obvious difficulties in such work, given the
complicated history of galaxies and the uncertainty in the IMF.  The
latter is of particular importance and gives rise to a relatively
large uncertainty in the estimates as we will discuss
below. Nevertheless, the dependence of stellar mass-to-light ratio
with color is fairly well constrained.

Bell \& de Jong (2001) used a suite of galaxy evolution models to show
that one expects substantial variation in stellar mass-to-light ratio
as a function of galaxy color. Although their work focussed on the
properties of spiral galaxies, comparison with results for early-type
galaxies suggest that we can extend their calculation to these
galaxies as well. Bell \& de Jong (2001) find the models are well
described by a linear relation between $\log M/L$ and $B-V$ color, and
provide tables with the slope and intercept of these relations.  Their
results, however, are for $z=0$, but we have converted their results
to the mean redshift of our lenses $(z=0.32)$ using predictions based
on the PEGASE code (Fioc \& Rocca-Volmerange 1997), provided by
D. LeBorgne.  Compared to $z=0$, the galaxies are slightly bluer, with
stellar mass-to-light ratios $\sim 25\%$ lower at $z=0.32$.

\begin{figure}
\begin{center}
\leavevmode
\hbox{%
\epsfxsize=9cm
\epsffile{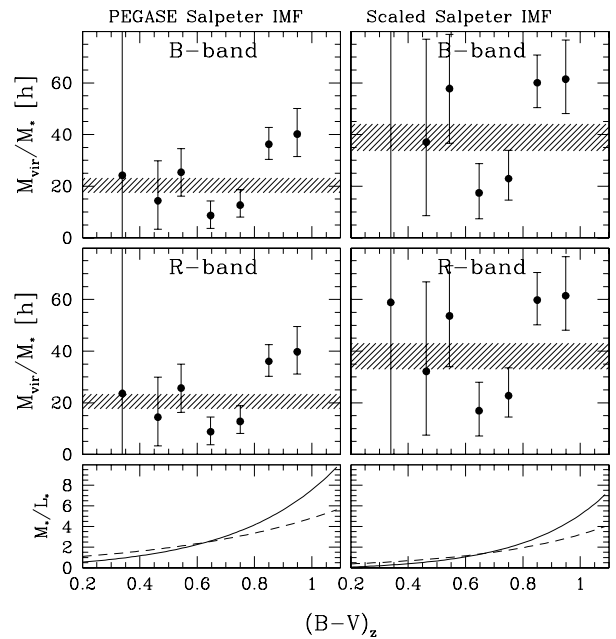}}
\caption{\footnotesize Lower panels show the stellar mass-to-light
ratios in the $B$-band (solid lines) and $R$-band (dashed lines) for
the results from PEGASE models using a Salpeter IMF and $Z=0.02$
(left) and a scaled Salpeter IMF (right) from Table~1 from Bell \& de
Jong (2001).  The mass-to-light ratios have been evolved to $z=0.32$,
which corresponds to the mean redshift of our lens sample. The upper
panels show the resulting ratios of virial mass and stellar mass for
the $B$ and $R$-band data. For a given IMF, the results obtained in
the different filters agree well, but it is clear that the mean values
(indicated by the shaded areas) depend strongly on the adopted IMF.
\label{fstar}}
\end{center}
\end{figure}

The resulting stellar mass-to-light ratio is most sensitive to the
assumed IMF and we will consider two `extreme' cases, such that our
results should bracket the real properties of galaxies. The first IMF
we consider is the one proposed by Salpeter (1955). The $z=0.32$
stellar mass-to-light ratios based on the PEGASE code by Fioc \&
Rocca-Volmerange (1997) are presented in the bottom left panel of
Figure~\ref{fstar} for the $B$ (solid line) and $R$-band (dashed
line). As noted by Bell \& de Jong (2001) a standard Salpeter (1955)
IMF results in mass-to-light ratios that are too high to fit rotation
curves of spiral galaxies.  Hence, this model can be considered
extreme in the sense that it provides a high estimate for the mass in
stars. Instead, Bell \& de Jong (2001) propose a scaled Salpeter IMF,
which is equivalent to reducing the number of low mass stars (which
contribute to the mass, but not to the luminosity). We use the
parameters from their Table~1.  The results for this IMF, which fits
the rotation curve data better, are shown in the bottom right panel of
Figure~\ref{fstar}. 

We use these model stellar mass-to-light ratios to calculate the ratio
$M_{\rm vir}/M_*$ as a function of color in both $B$ and $R$ band.
The results are also presented in Figure~\ref{fstar}. For a given
model, the results between the two filters agree very well, and the
average ratios are listed in Table~\ref{tab_star}, for a Hubble
parameter of $H_0=71$ km/s/Mpc. However, the two models yield
significantly different ratios. 

\begin{table*}
\begin{center}
\caption{Stellar mass and baryon fractions\label{tab_star}}
\begin{tabular}{llcclcc}
\hline
\hline
\multicolumn{2}{c}{} & \multicolumn{2}{c}{PEGASE} & ~~ & \multicolumn{2}{c}{SCALED}\\
 &                   & $B$ & $R$ & & $B$ & $R$ \\
\hline
all       & $M_{\rm vir}/M_*$           & $14\pm2$                  & $15\pm2$                  & & $27\pm4$                  & $28\pm4$                    \\
          & $f_{\rm bar\rightarrow *}$  & $0.41^{+0.07}_{-0.05}$    & $0.39^{+0.06}_{-0.05}$    & & $0.22^{+0.04}_{-0.03}$    & $0.21^{+0.03}_{-0.03}$      \\
          & $f^{\rm gal}_{\rm bar}$     & $0.070^{+0.012}_{-0.009}$ & $0.065^{+0.010}_{-0.008}$ & & $0.037^{+0.005}_{-0.004}$ & $0.035^{+0.005}_{-0.004}$   \\
          &                             &                           &                           & &                           &                             \\
$B-V<0.8$ & $M_{\rm vir}/M_*$           & $9\pm2$                   & $10\pm3$                  & & $17\pm5$                  & $18\pm5$                    \\
          & $f_{\rm bar\rightarrow *}$  & $0.65^{+0.20}_{-0.14}$    & $0.60^{+0.20}_{-0.12}$    & & $0.34^{+0.11}_{-0.08}$    & $0.32^{+0.12}_{-0.07}$      \\
          & $f_{\rm bar}$               & $0.11^{+0.03}_{-0.02}$    & $0.10^{+0.03}_{-0.02}$    & & $0.057^{+0.018}_{-0.013}$ & $0.055^{+0.021}_{-0.011}$   \\
          &                             &                           &                           & &                           &                             \\
$B-V>0.8$ & $M_{\rm vir}/M_*$           & $26\pm4$                  & $28\pm4$                   & & $42\pm6$                  & $45\pm6$                    \\
          & $f_{\rm bar\rightarrow *}$  & $0.22^{+0.03}_{-0.03}$     & $0.21^{+0.03}_{-0.03}$    & & $0.14^{+0.02}_{-0.02}$    & $0.13^{+0.02}_{-0.02}$      \\
          & $f^{\rm gal}_{\rm bar}$     & $0.038^{+0.006}_{-0.005}$ & $0.036^{+0.005}_{-0.004}$ & & $0.024^{+0.004}_{-0.003}$ & $0.022^{+0.004}_{-0.003}$   \\
          &                             &                 &               \\
\hline
\end{tabular}
\end{center}
\tablecomments{Note: Results for the PEGASE model using a standard
Salpeter IMF and scaled Salpeter IMF from Bell \& de Jong (2001).
These models have been evolved to a redshift of $z=0.32$ to allow for
a direct comparison with the measurements. For different color
selections, the rows list respectively, the ratio of virial mass over
stellar mass, the implied fraction of baryons transformed into stars
and the total visible baryon fraction in galaxies. Note that the
results for the $B$ and $R$-band are not independent. We have adopted
a Hubble constant of $H_0=71$km/s/Mpc and a universal baryon fraction
of $\Omega_b/\Omega_m=0.17$ (e.g., Spergel et al. 2003).}
\end{table*}

Observations of the cosmic microwave background (e.g., Spergel et
al. 2003) have yielded accurate measurements of the baryon fraction in
the universe. Based on WMAP observations, Spergel et al. (2003)
obtained $\Omega_b h^2=0.024\pm0.001$ and $\Omega_m h^2=0.14\pm 0.02$.
If we assume that baryons do not escape the dark matter overdensity
they are associated with, the ratio of mass in baryons to the total
mass of the halo is $M_{\rm bar}/M_{\rm vir}=\Omega_b/\Omega_m=
0.17\pm0.03$. In the following, we will also assume $H_0$=71/km/s/Mpc,
which is the currently favoured value.

For the PEGASE Salpeter model, this implies that the fraction of the
mass in stars is $0.070\pm0.011$ (average of the $B$ and $R$ value),
whereas the scaled Salpeter IMF yields a lower value of
$0.037\pm0.005$. Comparison with the value of $\Omega_b/\Omega_m$ from
CMB measurements suggests that only $\sim 40\%$ and $\sim 22\%$ of the
baryons are converted into stars for the standard and scaled Salpeter
IMFs respectively. The actual results for the two filters considered
here are indicated separately in Table~\ref{tab_star} by $f_{{\rm
bar}\rightarrow *}$.

Table~\ref{tab_star} also lists the average results when we consider
blue and red galaxies separately. The implied star formation
efficiencies for early-type galaxies are low. We note that similar
efficiencies have been inferred for galaxy clusters (e.g., Lin, Mohr
\& Stanford 2003).

Interestingly, our results imply that late-type galaxies convert a
$\sim 2$ times larger fraction of baryons into stars. This result is
robust, as it does not depend much on the adopted IMF. Guzik \& Seljak
(2002) also found a factor of $\sim 2$ difference in star formation
efficiency between early and late-type galaxies, in good agreement
with our findings. Hence, these results provide very important, direct
observational constraints on the relative star formation efficiency
during galaxy formation for different galaxy types.

These findings suggest that the mechanism for the formation of early
type galaxies is somehow more efficient in removing gas compared to
late type galaxies. Ram pressure stripping might be more prevalent,
given that early type galaxies are typically found in high density
regions, or they might form while developing strong winds that blow
out most of the baryons.

Irrespective of the process responsible for ejecting baryons, the
resulting galaxy will always have a stellar fraction which is greater
or equal to the fraction of stars in its progenitors. If we consider
the situation where early-type galaxies form through mergers, it is
clear that  not all early-type galaxies can be the result of
merging the late-type galaxies studied in this paper. Ejecting $\sim
60\%$ of the stars during the merger process might seem an option, but
this is hard to envision without removing a similar fraction of the
dark matter halo.

Hence, the progenitors of early-type galaxies must have had a low
fraction of their mass in stars.  This could be achieved if early-type
galaxies (or their progenitors) formed early on without forming new
stars at later times (because they lost their gas) and if later type
galaxies sustained their star formation for a much longer time, thus
building up a larger fraction of mass in stars. Recent estimates of
the star formation rates of high redshift galaxies, suggest a
qualitatively similar picture, in which early-type galaxies formed the
bulk of their stars very early on with a sharp drop in star formation
rates at $z\sim 2$, and less massive (late-type) galaxies continue to
form most of their stars at a later time and over a much longer period
of time (e.g., McCarthy et al. 2004; Juneau et al. 2005).

\subsection{Visible baryon fraction}

In addition to stars, galaxies contain some gas, which needs to be
included if we are to do a full accounting of the visible baryon
contents of galaxies. Although the amount of molecular hydrogen is
uncertain, the amount of neutral hydrogen is relatively well
determined from 21cm line studies. The relative amount of HI gas is a
function of galaxy type, with late-type galaxies being more gas
rich. We use the results from Roberts \& Haynes (1994) for the $M_{\rm
HI}/L_B$ ratio to estimate the amount of mass in gas (correcting the
hydrogen mass to account for the primoridial helium abundance). The
inclusion of gas slightly raises the mass in detected baryons for the
bluest galaxies but this component is negligible for the red galaxies.
Adding estimates for the amount of molecular hydrogen does not change
the numbers either.

The resulting fraction of the mass in baryons in galaxies, $f^{\rm
gal}_{\rm bar}$, is listed in Table~\ref{tab_star} as well. Only for
the blue galaxies, under the assumption of a standard Salpeter (1955)
IMF, is the baryon fraction marginally consistent with the value
determined from observations of the CMB (Spergel et
al. 2003). However, as discussed earlier, the results for this model
should be considered upper limits to the baryon fraction, given that
the stellar mass-to-light ratios are too high to fit rotation curves
(e.g., Bell \& de Jong 2001).  The results from the Scaled IMF from
Bell \& de Jong are probably more representative of the actual baryon
fractions in galaxies, thus implying that a significant fraction of
the gas must have been lost.

\section{Conclusions}

We have measured the weak lensing signal as a function of restframe
$B$, $V$, and $R$-band luminosity for a sample of `isolated' galaxies,
with photometric redshifts $0.2<z<0.4$. This selection of relatively
isolated galaxies minimizes the contribution of group/cluster halos
and nearby bright galaxies.

The photometric redshifts were derived by Hsieh et al. (2005) using
$BVR_Cz'$ photometry from the Red-Sequence Cluster Survey. To add to
the extensive study described in Hsieh et al. (2005), we confronted
the photometric redshifts to tests that are unique to weak
lensing. These results showed that the lensing signal around a sample
of foreground galaxies scales with source redshift as expected.  The
photometric redshift distribution determined by Hsieh et al. (2005)
suggests that the mean redshift of galaxies used in the measurement of
the lensing signal by large scale structure (Hoekstra et al. 2002a;
2002b) is somewhat higher than previously assumed. If correct, this
would imply a somewhat lower value for the normalization of the matter
power spectrum, $\sigma_8$, compared to the published results. The
difference is expected to be less than $\sim 10\%$, but with the
current data we cannot reliably quantify the size of the change.

Virial masses were determined by fitting an NFW model to the
tangential shear profile. Note, that intrinsic scatter in the
mass-luminosity relation will result in an underestimate of the mean
virial mass for a galaxy of a given luminosity, as suggested by
Tasitsiomi et al. (2004b). The magnitude of this effect depends on the
assumed scatter, and we have ignored this in our analysis. We found
that the virial mass as a function of luminosity is well described by
a power-law with a slope of $\sim 1.5$, with similar slopes for the
three filters considered here. This result agrees with other
observational studies (Guzik \& Seljak, 2002) and predictions from
semi-analytic models of galaxy formation (e.g., Kauffmann et al. 1999;
Benson et al. 2000; van den Bosch et al. 2003). For a galaxy with a
fiducial luminosity of $10^{10}h^{-2}$L$_{B\odot}$ we obtained a mass
$M_{\rm vir}=9.9^{+1.5}_{-1.3}\times 10^{11}$M$_\odot$. Converting
this result to match the filter and definition for the mass used by
Guzik \& Seljak (2002), yields a mass of $M_{200}=(11.7\pm1.7)\times
10^{10}h^{-1} \msun$ for a galaxy with a luminosity of $1.1\times
10^{10}h^{-2}\mlg$, in agreement with Guzik \& Seljak (2002), who
found $M_{200}=(9.3\pm1.6)\times 10^{11}h^{-1}\msun$.

We examined the efficiency with which baryons are converted into
stars. To do so, we used the restframe $B-V$ color as a measure of the
mean stellar mass-to-light ratio. The color also provides a crude
indicator of galaxy type (e.g., Robert \& Haynes 1994). We considered
a standard and a scaled Salpeter IMF (see Bell \& de Jong 2001). The
latter is more realistic, whereas the former yields stellar
mass-to-light ratios that are too high to fit rotation curves of
spiral galaxies.

Irrespective of the adopted IMF, we found that the stellar mass
fraction is about a factor of two lower for early-type galaxies, as
compared to late-type galaxies. Including the fraction of baryons in
gas only increases the fraction of observed baryons slightly. Hence,
our results suggest that galaxy formation is very inefficient in
turning baryons into stars and in retaining baryons. These results
provide important, direct observational constraints for models of
galaxy formation.

Under the assumption that the scaled Salpeter IMF is correct, our
results imply that late-type galaxies convert $\sim 33$\% of baryons
into stars. Early-type galaxies do much worse, with an efficiency of
$\sim 14$\%. This implies that the progenitors of early-type galaxies
have a low fraction of their mass in stars.  A possible explanation of
this result is that early-type galaxies formed early on and stopped
forming new stars, because they lost most of their baryons (e.g.,
through winds or ram pressure stripping).  If later type galaxies, on
the other hand, continued to form stars this would lead to a higher
stellar mass fraction. Such a scenario is, at least qualitatively, in
agreement with recent estimates of the star formation rates of high
redshift galaxies (e.g., McCarthy et al. 2004; Juneau et al. 2005).


\begin{thebibliography}{}

\bibitem{Bartelmann96}
Bartelmann, M. 1996, A\&A, 313, 697

\bibitem{Bell01}
Bell, E.F. \& de Jong, R.S 2001, ApJ, 550, 212

\bibitem{Benson00}
Benson, A.J., Cole, S., Frenk, C.S., Baugh, C.M., \& Lacey, C.G. 
2000, MNRAS, 311, 793

\bibitem{BN02}
Bernstein, G.M. \& Norberg, P. 2002, AJ, 124, 733

\bibitem{Bryan98}
Bryan, G.L. \& Norman, M.L. 1998, ApJ, 495, 80

\bibitem{Bullock01}
Bullock, J.S. et al. 2001, MNRAS, 321, 559

\bibitem{Capak}
Capak, P. et al. 2004, AJ, 127, 180

\bibitem{Connolly}
Connolly, A.J., Csabai, I. \& Szalay, A.S. 1995, AJ, 110, 2655

\bibitem{Cowie} 
Cowie, L.L., Barger, A.J., Hu, E.M., Capak, P., \& Songaila, A. 2004,
AJ, 127, 3137

\bibitem{Diemand}
Diemand, J., Moore, B. \& Stadel, J. 2004, MNRAS, 353 624

\bibitem{Dubinski91}
Dubinski, J. \& Carlberg, R.G. 1991, ApJ, 378, 496

\bibitem{FJ76}
Faber, S.M. \& Jackson, R.E. 1976, ApJ, 204, 668

\bibitem{pegase}
Fioc, M. \& Rocca-Volmerange, B. 1997, A\&A, 326, 950
\bibitem{FISCH00}
Fischer, P., et al. 2000, AJ, 120, 1198

\bibitem{Fukugita} 
Fukugita, M., Ichikawa, T., Gunn, J.E., Doi, M., Shimasaku, K., \&
Schneider, D.P. 1996, AJ, 111, 1748

\bibitem{GLAD04}
Gladders, M.D. \& Yee, H.K.C. 2005, ApJS, 157, 1

\bibitem{GUZ01}
Guzik, J., \& Seljak, U. 2001, MNRAS, 321, 439
 
\bibitem{GUZ02}
Guzik, J., \& Seljak, U. 2002, MNRAS, 335, 311

\bibitem{Hayashi}
Hayashi, E., Navarro, J.F., Power, C., Jenkins, A., Frenk, C.S.,
White, S.D.M., Springel, V., Stadel, J., \& Quinn, T.R. 2004, MNRAS, 355, 794

\bibitem{HOEK02a}
Hoekstra, H., Yee, H.K.C., Gladders, M.D., Barrientos, L.F., Hall,
P.B., \& Infante, L. 2002a, ApJ, 572, 55

\bibitem{HOEK02b}
Hoekstra, H., Yee, H.K.C., \& Gladders, M.D. 2002b, ApJ, 577, 595

\bibitem{HOEK02c}
Hoekstra, H., van Waerbeke, L., Gladders, M.D., Mellier, Y., \& Yee,
H.K.C. 2002c, ApJ, 577, 604

\bibitem{HOEK02d}
Hoekstra, H., Franx, M., Kuijken, K. \& van Dokkum, P.G. 2002d, 
MNRAS, 333, 911

\bibitem{HOEK04}
Hoekstra, H., Yee, H.K.C., \& Gladders, M.D. 2004, ApJ, 606, 67

\bibitem{HSIEH04}
Hsieh, B.C., Yee, H.K.C., Lin, H., \& Gladders, M.D. 2005, ApJS, in press

\bibitem{HUD98}
Hudson, M.J., Gwyn, S.D.J., Dahle, H., \& Kaiser, N. 1998, ApJ, 503, 531

\bibitem{J05}
Juneau, S. et al. 2005, ApJ, 619, L135

\bibitem{k99a}
Kauffmann, G., Colberg, J.M., Diaferio, A., White, S.D.M.	
1999, MNRAS, 303, 188

\bibitem{Keeton98}
Keeton, C.R., Kochanek, C.S., \& Falco, E.E. 1998, ApJ, 509, 561

\bibitem{Lin99}
Lin, H., Yee, H.K.C., Carlberg, R.G., Morris, S.L., Sawicki, M., 
Patton, D.R., Wirth, G. \& Shepherd, C.W. 1999, ApJ, 518, 533

\bibitem{Lin2003}
Lin, Y.-T., Mohr, J.J., \& Stanford, S.A. 2003, ApJ, 591, 749

\bibitem{McCarthy04}
McCarthy, P.J. et al. 2004, ApJ, 614, L9

\bibitem{McKay01}
McKay, T.A., et al. 2001, ApJ, submitted, astro-ph/0108013

\bibitem{Moore99}
Moore, B., Quinn, T., Governato, F., Stadel, J., \& Lake, G. 1999,
MNRAS, 310, 1147

\bibitem{NFW95}
Navarro, J.F., Frenk, C.S., \& White, S.D.M. 1995, MNRAS, 275, 56
 
\bibitem{NFW96}
Navarro, J.F., Frenk, C.S., \& White, S.D.M. 1996, ApJ, 462, 563
 
\bibitem{NFW97}
Navarro, J.F., Frenk, C.S., \& White, S.D.M. 1997, ApJ, 490, 493

\bibitem{Parker}
Parker, L.C., Hudson, M.J., Carlberg, R.G., \& Hoekstra, H. 2005, ApJ, submitted

\bibitem{Roberts94}
Roberts, M.S., \& Haynes, M.P. 1994, ARA\&A, 32, 115

\bibitem{Salpeter}
Salpeter, E.E. 1955, ApJ, 121, 61

\bibitem{Sheldon}
Sheldon, E.S., et al. 2004, AJ, 127, 2544

\bibitem{Smith01}
Smith, D.R., Bernstein, G.M., Fischer, P., \& Jarvis, M. 2001, ApJ,
551, 643

\bibitem{Spergel2003}
Spergel, D.N. et al. 2003, ApJS, 148, 175

\bibitem{Tasitsiomi1}
Tasitsiomi, A., Kravtsov, A.V., Gottl{\"o}ber, S. \& Klypin, A.A. 2004a, 
ApJ, 607, 125

\bibitem{Tasitsiomi2} 
Tasitsiomi, A., Kravtsov, A.V., Wechsler, R.H. \& Primack, J.R. 2004b, 
ApJ, 614, 533

\bibitem{TF77}
Tully, R.B. \& Fisher, J.R. 1977, A\&A, 54, 661

\bibitem{Albada}
van Albada, T.S., \& Sancisi, R. 1986,  Phil. Trans. Roy. Soc.,
London, A320, 447

\bibitem{vdB}
van den Bosch, F., Yang, X., \& Mo, H.J. 2003, MNRAS, 340, 771

\bibitem{vD96}
van Dokkum, P.G., \& Franx, M. 1996, MNRAS, 281, 985

\bibitem{Verheijen}
Verheijen, M.A.W. 2001, ApJ, 563, 694

\bibitem{Wilson01}
Wilson, G., Kaiser, N., Luppino, G.A., Cowie, L.L. 2001, ApJ, 555, 572

\bibitem{Wirth}
Wirth, G.D. et al. 2004, AJ, 127, 3121

\bibitem{WB00}
Wright, C.O. \& Brainerd, T.G. 2000, ApJ, 534, 34

\bibitem{Yee00}
Yee, H.K.C. et al. 2000, ApJS, 129, 475

\end{thebibliography}
\end{document}